\documentclass[prd,twocolumn,superscriptaddress,floatfix,amsmath,amssymb,amsfonts,longbibliography,nofootinbib]{revtex4-1}

\usepackage{float} 

\usepackage{comment}
\usepackage[normalem]{ulem}
\usepackage[english]{babel}
\usepackage{graphicx}
\usepackage{dcolumn}
\usepackage{bm}
\usepackage{blindtext}
\usepackage{verbatim}
\usepackage{mathrsfs}
\usepackage{musicography}
\usepackage{amsmath}
\usepackage{cancel}
\usepackage{physics}
\usepackage{epstopdf}
\usepackage{mathtools}
\usepackage{tensor}
\usepackage{color}
\usepackage[usenames,dvipsnames]{pstricks}
\usepackage{epsfig}
\usepackage{pst-grad} 
\usepackage{pst-plot} 
\usepackage{hyperref}
\usepackage{verbatim}
\usepackage{slashed}

\newcommand{\mf}{\mathsf}

\newcommand{\trr}[1]{\textcolor{red}{#1}}

\renewcommand{\arcsin}{\cos\!{}^{-1}\!\!}

\allowdisplaybreaks[1] 

\begin{document}

\title{Spacetime curvature from ultra rapid measurements of quantum fields}

\author{T. Rick Perche}
\email{trickperche@perimeterinstitute.ca}
\affiliation{Department of Applied Mathematics, University of Waterloo, Waterloo, Ontario, N2L 3G1, Canada}
\affiliation{Perimeter Institute for Theoretical Physics, Waterloo, Ontario, N2L 2Y5, Canada}
\affiliation{Institute for Quantum Computing, University of Waterloo, Waterloo, Ontario, N2L 3G1, Canada}

\author{Ahmed Shalabi}
\email{ashalabi@uwaterloo.ca}
\affiliation{Department of Physics and Astronomy, University of Waterloo, Waterloo, Ontario, N2L 3G1, Canada}

\begin{abstract}
    We write the curvature of spacetime in terms of the excitation probability of  particle detectors ultra-rapidly coupled to a quantum field. More precisely, we provide an expansion for the excitation probability of a smeared UDW detector delta-coupled to a real scalar quantum field in a curved background. Using a short distance expansion for the Wightman function, we express the excitation probability of a detector as the transition probability in Minkowski spacetime plus correction terms written as a function of the curvature tensors and the detector size. Comparing the excitation probability in curved spacetimes with its flat analog, we are able to express the components of the Ricci and Riemann curvature tensors as a function of physically measurable excitation probabilities of different shaped detectors.
\end{abstract}

\maketitle

\section{Introduction}

At its core, general relativity is a theory of gravity phrased operationally in terms of measurements of distances and time using classical rulers and clocks. Quantizing these notions has been a major problem of theoretical physics for the past century and, as of today, there is still no complete theory of quantum gravity. Nevertheless, there are multiple effective tools that can be used in order to better understand the relationship of gravity and quantum physics in low energy regimes. In particular, the behaviour of quantum fields in curved spacetimes can be well described using a semiclassical theory, where the background is classical and the matter fields are quantum. Although this approach does not provide a full theory of quantum gravity, it gives important results, such as the Unruh and Hawking effects~\cite{fullingUnruhEffect,HawkingRadiation,Unruh1976,Unruh-Wald} and the model of inflation~\cite{inflation}, which describes the universe fractions of seconds after its creation. A more recent application of this semiclassical theory is to rephrase classical notions of space and time intervals in terms of properties of quantum fields~\cite{achim,achimQGInfoCMB,achim2,mygeometry}. As argued in~\cite{achim2,mygeometry}, this rephrasing might lead to a quantum theory of spacetime, which could redefine the notions of distance and time close to the Planck scale.

In order to relate the spacetime geometry with properties of a quantum field theory (QFT), it is necessary to study the specific way that the background spacetime affects a quantum field. The effect of curvature in the correlation function of a QFT has been thoroughly studied in the literature~\cite{Wald2,kayWald,achim}. 
In fact, it is possible to show that, within short scales, the behaviour of the correlations of a quantum field can be written as the correlations in flat spacetime added to terms that involve corrections due to curvature~\cite{birrell_davies,DeWittExpansion}. This suggests that if one finds a mechanism to locally probe these correlations, one would then be able to recover the geometry of spacetime. 

One way of probing quantum fields locally, and recovering their correlation functions, is through the use of particle detector models~\cite{pipo,mygeometry}. Broadly speaking, particle detector models are localized non-relativistic quantum systems that couple to a quantum field. Examples of physical realizations of these range from atoms probing the electromagnetic field~\cite{Pozas2016,Nicho1,richard} to nucleons interacting with the neutrino fields~\cite{neutrinos,antiparticles,fermionicharvesting}. After their first introduction by Unruh and DeWitt in~\cite{Unruh1976,DeWitt}, these models found many different uses for studying a wide range of phenomena of quantum field theories in both flat and curved spacetimes. There are several applications of these models, such as the study of the entanglement in quantum fields~\cite{Valentini1991,Reznik1,reznik2,Retzker2005,Pozas-Kerstjens:2015,Pozas2016}, the Unruh effect~\cite{Unruh-Wald,HawkingRadiation,fullingHadamard,matsasUnruh,mine} and Hawking radiation~\cite{HawkingRadiation,HawkingMain,WaldRadiation,detRadiation,detectorsCavitiesFalling}. Moreover, particle detectors can be used to  provide a measurement framework for quantum field theory~\cite{FVdebunked,chicken}, to probe the topology~\cite{topology} and geometry of spacetime~\cite{mygeometry}, among other applications~\cite{pipo,teleportation,adam}. 

In this manuscript we show how it is in principle possible to recover the curvature of spacetime using smeared particle detectors ultra rapidly coupled to a quantum field~\cite{deltaCoupled,nogo}. Smeared particle detectors have a finite spatial extension, which can be controlled to probe the quantum field in different directions. The effect of the spacetime curvature in the correlation function of the quantum field then affects the transition probabilities of the detector. We precisely quantify how curvature affects the response of particle detectors, so that by comparing the response of a detector in curved spacetimes with what would be seen in Minkowski, one can infer the spacetime curvature. Using particle detectors with different shapes then gives access to the spacetime curvature in different directions, so that it is possible to reconstruct the full Riemann tensor at the center of the detector's trajectory, and all geometrical quantities derived from it.

Our results are another instance of rephrasing geometrical properties of spacetime in terms of measurements of observable quantities of quantum fields~\cite{achim,achimQGInfoCMB,achim2,mygeometry}. We argue that such rephrasing is an important step towards understanding the relationship between quantum theory and gravity. This sets the grounds for future works which might provide a detailed answer to how to define the notions of space and time in scales where the classical notions provided by general relativity fail to work.

This paper is organized as follows. 
In Section \ref{sec:detectors} we describe the coupling and the dynamics of a particle detector ultra rapidly coupled to a massless scalar field. In Section \ref{sec:curvature} we write the excitation probability of an ultra rapidly coupled particle detector as an expansion in the detector size, with coefficients related to the spacetime curvature. In Section \ref{sec:recovery} we provide a protocol such that one can recover the spacetime curvature from the excitation probability of particle detectors. The conclusions of our work can be found in Section \ref{sec:conclusion}.


\section{Ultra rapid sampling of quantum fields}\label{sec:detectors}

In this section we describe the particle detector model that will be used in this manuscript. We consider a two-level Unruh-DeWitt (UDW) detector model coupled to a free massless scalar quantum field $\hat{\phi}(\mf x)$ in a \mbox{$D = n+1$} dimensional spacetime $\mathcal{M}$ with metric $g$. The Lagrangian associated with the field can be written as
\begin{equation}
    \mathcal{L} = - \frac{1}{2} \nabla_\mu \phi \nabla^\mu \phi,
\end{equation}
where $\nabla$ is the Levi-Civita connection. We will not be concerned with the details of the quantization of the field here. However, we will assume that the state of the field is a Hadamard state, for reasons that will become clear in Section \ref{sec:curvature}. Moreover, Hadamard states are those for which it is possible to associate a finite value to the stress-energy tensor of the quantum field~\cite{birrell_davies,fewsterNecessityHadamard}, which makes them appealing from a physical perspective.

The detector is modelled by a two-level system undergoing a timelike trajectory $\mf z(\tau)$ in $\mathcal{M}$ with four-velocity $u^\mu(\tau)$ and proper time parameter $\tau$. We pick Fermi normal coordinates $(\tau,\bm{x})$ around $\mf z(\tau)$ (for more details we refer the reader to~\cite{poisson,us,us2}). We assume the proper energy gap of the two-level system to be $\Omega$, such that its free Hamiltonian in its proper frame is given by \mbox{$\hat{H}_D = \Omega \hat{\sigma}^+\hat{\sigma}^-$}, where $\hat{\sigma}^\pm$ are the standard raising/lowering ladder operators. The interaction with the  scalar field is prescribed by the scalar interaction Hamiltonian density
\begin{equation}
    \hat{h}_I(\mf x) = \tilde{\lambda} \Lambda(\mf x) \hat{\mu}(\tau)\hat{\phi}(\mf x),
\end{equation}
where  \mbox{$\hat{\mu}(\tau) = e^{- i \Omega \tau} \hat{\sigma}^- +e^{i \Omega \tau} \hat{\sigma}^+ $} is detector monopole moment, $\tilde{\lambda}$ is the coupling constant, and $\Lambda(\mf x)$ is a scalar function that defines the spacetime profile of the interaction. This setup defines the interaction of a UDW detector with a real scalar quantum field, and has been thoroughly studied in the literature~\cite{birrell_davies,pipo,mygeometry,antiparticles,fermionicharvesting,Unruh1976,DeWitt,Pozas-Kerstjens:2015,us,us2}. This model also has a physical appeal, as it has been shown to reproduce realistic models, such as atoms interacting with the electromagnetic field~\cite{Pozas2016,Nicho1,richard} and nucleons with the neutrino fields~\cite{neutrinos,antiparticles,fermionicharvesting}.

Under the assumption that the shape of the interaction between the detector and the field is constant in the detector's frame i.e. a rigid detector, we can write the spacetime smearing function as $\Lambda(\mf x) = \chi(\tau)f(\bm x)$, where now $f(\bm x)$ (the smearing function) defines the shape of the interaction and $\chi(\tau)$ (the switching function) controls the strength and the duration of the coupling. This decomposition also allows one to control the proper time duration of the interaction by considering $\chi(\tau) = \eta\, \varphi(\tau/T)/T$ for a positive compactly supported function $\varphi$ that is $L^1(\mathbb{R})$ normalized and symmetric with respect to the origin. Here $\eta$ and $T$ are parameters with units of time, which ensure that $\chi(\tau)$ is dimensionless. In this manuscript we will be particularly interested in an ultra rapid coupling~\cite{deltaCoupled,nogo}, which is obtained when $T\longrightarrow 0$ and $\chi(\tau) \longrightarrow \eta\, \delta(\tau)$.

The evolution of the system after an interaction is implemented by the time evolution operator
\begin{equation}\label{U1}
    \hat{U} = \mathcal{T}_\tau \exp\left(-i \int_\mathcal{M} \dd V \hat{h}_I(\mf x)\right),
\end{equation}
where $\mathcal{T}_\tau$ denotes time ordering with respect to $\tau$ and $\dd V = \sqrt{-g} \,\dd^{D}x$ is the invariant spacetime volume element. In the case of ultra rapid coupling, where \mbox{$\chi(\tau) = \eta \, \delta(\tau)$}, Eq. \eqref{U1} simplifies to
\begin{align}
    \hat{U} &=  e^{-i \hat{\mu}\, \hat{Y}(f)} = \text{cos}(\hat{Y}) -i \hat{\mu}\, \text{sin}(\hat{Y}),
\end{align}
where
\begin{align}
        \hat{Y}(f)&= \lambda \int \dd^3\bm x \sqrt{-g} f(\bm x) \hat{\phi}(\bm x),
\end{align}
with $\hat{\mu} = \hat{\mu}(0)= \hat{\sigma}^+ + \hat{\sigma}^- $, $\lambda =\tilde{\lambda}\eta$ and \mbox{$\hat{\phi}(\bm x) = \hat{\phi}(\tau = 0,\bm x)$} is the field evaluated at the rest space associated to the interaction time, $\tau = 0$. In the equation above the integral is performed over the rest space of the system at $\tau = 0$.

We consider a setup where the detector starts in the ground state $\hat{\rho}_{\textrm{d},0}= \hat{\sigma}^-\hat{\sigma}^{+}$ and the field starts in a given Hadamard state $\omega$
. The final state of the detector after the interaction, $\hat{\rho}_{\textrm{d}}$, will then be given by the partial trace over the field degrees of freedom. It can be written as
\begin{align}
    \hat{\rho}_{\textrm{d}} =& \omega(\hat{U}\hat{\rho}_{\textrm{d},0} \hat{U}^\dagger) \nonumber\\
    =& \omega(\cos^2(\hat{Y})) \hat{\sigma}^-\hat{\sigma}^+ + \omega(\sin^2(\hat{Y})) \hat{\sigma}^+ \hat{\sigma}^-\nonumber\\
    =& \frac{1}{2}\left(\openone + \omega(\cos{}(2\hat{Y}))\hat{\sigma}_z\right),
\end{align}
where we used $\hat{\mu} \hat{\rho}_{\text{d},0} \hat{\mu} = \hat{\sigma}^+ \hat{\sigma}^-$ and $\omega(\text{sin}(\hat{Y})\text{cos}(\hat{Y})) = 0$ due to the fact that this operator is odd in the field $\hat{\phi}$, and $\omega$ is a Hadamard state, so that it is quasifree~\cite{kayWald} and all of its odd point-functions vanish. In particular, notice that $\tr(\hat{\rho}_{\textrm{d}}) = \omega(\cos^2(\hat{Y})+\sin^2(\hat{Y})) = 1$, as expected. The excitation probability of the detector is then given by
\begin{equation}
    P = \tr(\hat{\rho}_{\text{d}}\sigma^+ \sigma^-) = \omega(\sin^2(\hat{Y})) = \frac{1}{2}\left(1 - \omega(e^{2 i \hat{Y}})\right),
\end{equation}
where we used $\sin^2\theta = \frac{1}{2}\left(1 - \cos(2\theta)\right)$ and the fact that $\omega(\cos{}(2\hat{Y})) = \omega(\text{exp}({2 i \hat{Y}}))$, because only the even part of the exponential contributes.

Moreover, there is a simple expression for the excitation probability in terms of a smeared integral of the field's Wightman function $W(\mf x,\mf x') = \omega(\hat{\phi}(\mf x)\hat{\phi}(\mf x'))$. Let
\begin{equation}\label{eq:L}
    \mathcal{L} = \lambda^2 \int \dd^n\bm x \dd^n \bm x'\,\sqrt{-g}\sqrt{-g'}\, f(\bm x) f(\bm x') W(\bm x,\bm x'),
\end{equation}
where $W(\bm x, \bm x') = W(\tau\! =\! 0, \bm x,\tau'\!=\! 0, \bm x')$. Then, we show in Appendix \ref{app:L} that if $\omega$ is a quasifree state, \mbox{$\omega(e^{2 i \hat{Y}}) = e^{-2 \mathcal{L}}$}, so that the excitation probability of the delta coupled detector can be written as
\begin{equation}\label{eq:P}
    P = \frac{1}{2}\left(1 - e^{-2 \mathcal{L}}\right).
\end{equation}
Notice that in the pointlike limit the detector is essentially sampling the field correlator at a single point. In this case, $\hat{\rho}_{\text{d}}\rightarrow \frac{1}{2}\openone$ and no information about the quantum field can be obtained. By considering finite sized detectors, it is then possible to sample the field in local regions, allowing one to recover information about both the field and its background spacetime. 

\section{The effect of curvature on the excitation probability}\label{sec:curvature}

In this section we will derive an expansion for the excitation probability of a particle detector rapidly interacting with a quantum field in curved spacetimes. From now on, we will focus in the case of  (3+1) dimensions. Our expansion will relate $P$ in Eq. \eqref{eq:P} with the excitation probability of a delta-coupled particle detector in Minkowski spacetime. By comparing these results, we will later be able to rewrite the components of the Riemann curvature tensor as a function of the excitation probability of the detector.

Notice that the detector's excitation probability in Eq. \eqref{eq:P} is entirely determined by $\mathcal{L}$ in \mbox{Eq. \eqref{eq:L}}, so that in order to obtain an expansion for the excitation probability, it is enough to expand $\mathcal{L}$. The first step in order to perform our expansion is to write the Wightman function in curved spacetimes as its flat spacetime analog added to an expansion in terms of curvature. Assuming the field state $\omega$ to be a Hadamard state, it can be shown that the correlation function of a quantum field can be written \mbox{as~\cite{fullingHadamard,fullingHadamard2,kayWald,equivalenceHadamard,fewsterNecessityHadamard}}
\begin{align}
    W(\mf x,\mf x') \!=\! \frac{1}{8\pi^2}\! \frac{\Delta^{1/2}( \mf x,\mf x')}{\sigma( \mf x, \mf x')} \!+\! v( \mf x,\mf x') \text{ln}|\sigma(\mf x, \mf x')|\!+\!w(\mf x,\mf x'),\label{Whada}
\end{align}
where $v(\mf x,\mf x')$ and $w(\mf x,\mf x')$ are regular functions in the limit $\mf x'\rightarrow \mf x$, $\Delta(\mf x,\mf x')$ is the Van-Vleck determinant (see~\cite{poisson}) and $\sigma(\mf x,\mf x')$ is Synge's world function, corresponding to one half the geodesic separation between the events $\mf x$ and $\mf x'$. In Eq. \eqref{Whada}, the function $w(\mf x,\mf x')$ contains the state dependence, while $v(\mf x,\mf x')$ is fully determined by the properties of both the field and the spacetime. We can then write
\begin{align}
    W(\mf x,\mf x') = \frac{1}{8\pi^2\sigma} \bigg[\Delta^{1/2} \!+ \! 8 \pi^2v_0\,\sigma\,&\text{ln}|\sigma|+8\pi^2 w_0\,\sigma \label{Wexpsigma}\\
    &\:\:\:\:\:\:\:\:\:\:+ \mathcal{O}(\sigma^2\,\text{ln}|\sigma|)\bigg],\nonumber
\end{align}
where $v_0 = v_0(\mf x,\mf x')$ and $w_0 = w_0(\mf x,\mf x')$ are the first order terms of an expansion of $v$ and $w$ in powers of $\sigma$~\cite{DeWittExpansion}. Notice that we have factored the Minkowski spacetime Wightman function for a massless field, $W_0(\mf x,\mf x') = \frac{1}{8\pi^2\sigma}$ in Eq. \eqref{Wexpsigma}. In~\cite{DeWittExpansion,poisson} it was shown that for a massless field, $v_0(\mf x,\mf x') = R(\mf x)/6 + \mathcal{O}(\sqrt{\sigma})$, so that the leading order contribution for the expansion is given by the Ricci scalar. The same is true for the state dependent part of the Wightman function, $w(\mf x,\mf x') = \omega_0(\mf x) + \mathcal{O}(\sqrt{\sigma})$ for a given function $\omega_0(\mf x)$ which determines the state contribution to $W(\mf x,\mf x')$ to leading order in $\sigma$. Moreover, the Van-Vleck determinant admits the following expansion
\begin{equation}
    \Delta^{\frac{1}{2}}(\mf x,\mf x') = 1 + \frac{1}{12}R_{\alpha \beta}(\mf x) \sigma^\alpha(\mf x,\mf x') \sigma^\beta(\mf x,\mf x'),
\end{equation}
where $\sigma^{\alpha}(\mf x,\mf x')$ denotes the tangent vector to the geodesic that connects $\mf x$ and $\mf x'$ such that its length corresponds to the spacetime separation between $\mf x$ and $\mf x'$. $\sigma_\alpha$ also corresponds to $\partial_\alpha \sigma$~\cite{poisson}.

Combining the results above, we find that the Wightman function of a quantum field in a Hadamard state can be approximated as
\begin{align}
    W(\mf x,\mf x') \approx W_0(\mf x,\mf x')\Big(1 +& \frac{1}{12}R_{\alpha\beta}(\mf x)\sigma^\alpha(\mf x,\mf x') \sigma^\beta(\mf x,\mf x') \nonumber\\
    &\!\!\!+\frac{4\pi^2}{3}R(\mf x)\,\sigma(\mf x,\mf x')\,\text{ln}|\sigma(\mf x,\mf x')|\nonumber\\
    &\:\:\:\:\:\:+8 \pi^2\omega_0(\mf x) \sigma(\mf x,\mf x')\Big),\label{WW0partial}
\end{align}
where $W_0(\mf x,\mf x')$ is the Wightman function in Minkowski spacetime. 

Equation \eqref{WW0partial} allows one to locally relate the Wightman function in curved spacetimes with its Minkowski counterpart. However, we wish to have an expansion in terms of the proper distance from the center of the interaction, $\mf z$. This proper distance can be expressed naturally in terms of Synge's world function in Fermi normal coordinates due to the fact that $x^i = \sigma^i(\mf z,\mf x)$, so $\sigma(\mf x,\mf x') = \frac{1}{2}\sigma^i(\mf z,\mf x)\sigma_i(\mf z,\mf x) =  \frac{1}{2}x^ix_i$. Thus, considering $\mf x$ and $\mf x'$ sufficiently close to the point $\mf z$, we can use the following approximations
\begin{align}
    \sigma(\mf x,\mf x') &\approx \sigma(\mf z,\mf x) - \sigma_\alpha(\mf z,\mf x) \sigma^\alpha(\mf z,\mf x') + \sigma(\mf z,\mf x')\nonumber,\\
    \sigma^\alpha(\mf x',\mf x) &\approx \sigma^\alpha(\mf z,\mf x) - \sigma^\alpha(\mf z,\mf x') = (\mf x-\mf x')^\alpha.
\end{align}
It is also possible to expand the Ricci scalar and the Ricci tensor according to $R(\mf x) = R(\mf z) + \mathcal{O}(r)$ and \mbox{$R_{\alpha\beta}(\mf x) = R_{\alpha\beta}(\mf z) + \mathcal{O}(r)$}, where $\mathcal{O}(r)$ denotes terms of order $r = \sqrt{x^i x_i}$. Analogously, we can expand the state dependent term as $\omega_0(\mf x) = \omega_0(\mf z)+\mathcal{O}(r)$. In the end we obtain an expression that relates $W(\mf x,\mf x')$ with $W_0(\mf x,\mf x')$, tensors evaluated at $\mf z$, and the effective separation vector between $\mf z$ and $\mf x$/$\mf x'$:
\begin{align}
    W(\mf x,\mf x') \approx W_0(\mf x,\mf x')\!\bigg[1&\! + \!\frac{1}{12}R_{ij}(x-x')^i(x-x')^j \label{WW0} \\
    &\!\!\!+\frac{2\pi^2}{3}R\,(\mf x-\mf x')^2\,\text{ln}\left|\tfrac{1}{2}(\mf x-\mf x')^2\right|\nonumber\\
    &\:\:\:\:\:\:\:\:\:\:\:\:\:\:\:\:\:\:\:\:+4\pi^2\omega_0(\mf z)(\mf x-\mf x')^2 \bigg],\nonumber
\end{align}
where the curvature tensors are all evaluated at the center point of the interaction, $\mf z$. In Eq. \eqref{WW0}, \mbox{$(x-x')^i = x^i - x'{}^i$} denotes the difference in Fermi normal coordinates of the points $\mf x$ and $\mf x'$ and \mbox{$(\mf x-\mf x')^2 = (x-x')^i(x-x')_i = r^2$}.

The last factor that was not yet considered in our expansion is the factor of $\sqrt{-g}$ terms that show up in the definition of $\mathcal{L}$ in Eq. \eqref{eq:L}. If the detector size is small enough compared to the radius of curvature of spacetime, we can employ the expansion of the metric determinant detailed in Appendix \ref{ap:fermi} around the center of the interaction $\mf z$. We then have
\begin{equation}\label{sqrtg}
    \sqrt{-g} = 1 + a_i x^i +\tfrac{1}{2}M_{ij} x^i x^j + \mathcal{O}(r^3),
\end{equation}
where $r = \sqrt{\delta_{ij}x^i x^j}$ corresponds to the proper distance from a point to $\mf z(0)$, $a_i$ is the four-acceleration of the trajectory at $\tau = 0$ and the tensor $M_{ij}$ is evaluated at $\mf z$. This tensor is explicitly given by
\begin{equation}
    M_{ij} = \tfrac{2}{3}R_{\tau i \tau j} - \tfrac{1}{3} R_{ij}.
\end{equation}



At this stage, we have all the tools required  to expand the excitation probability. Combining the results of Eqs. \eqref{WW0} and \eqref{sqrtg}, we can write the excitation probability of a smeared delta-coupled particle detector in a curved spacetime as the following short scale expansion
\begin{widetext}
\begin{align}\label{PcsPflat}
    P \approx P_0 + e^{-2 \mathcal{L}_0}\left( M_{ij}\mathcal{Q}^{ij}+2a_i\mathcal{D}^i+\frac{1}{12}R_{ij}\mathcal{L}^{ij} +\frac{2\pi^2}{3} R\, \mathcal{L}_R +4 \pi^2 \omega_0 \mathcal{L}_{\omega}\right),
\end{align}
\end{widetext}
where $P_0 = \frac{1}{2}\left(1 -e^{-2 \mathcal{L}_0}\right)$ and we have defined:
\begin{align}
    \mathcal{L}_0 \!&=\! \lambda^2 \!\!\!\int\!\!\dd^3 \bm x \dd^3 \bm x' \!f(\bm x) f(\bm x') W_0(\bm x,\bm x'),\nonumber\\
    \mathcal{Q}^{ij} \!&= \!\lambda^2  \!\!\!\int\!\!\dd^3 \bm x \dd^3 \bm x' \!f(\bm x) f(\bm x') W_0(\bm x,\bm x')x^ix^j,\nonumber\\
    \mathcal{D}^{i} \!&= \!\lambda^2  \!\!\!\int\!\!\dd^3 \bm x \dd^3 \bm x' \!f(\bm x) f(\bm x') W_0(\bm x,\bm x')x^i,\nonumber\\
    \mathcal{L}^{ij} \!&= \!\lambda^2  \!\!\!\int\!\!\dd^3 \bm x \dd^3 \bm x' \!f(\bm x) f(\bm x') W_0(\bm x,\bm x')(x\!-\!x')^i(x\!-\!x')^j,\nonumber\\
    \mathcal{L}_R \!&= \!\lambda^2 \!\!\! \int\!\!\dd^3 \bm x \dd^3 \bm x' \!f(\bm x) f(\bm x') W_0(\bm x,\bm x')\nonumber\\
    &\:\:\:\:\:\:\:\:\:\:\:\:\:\:\:\:\:\:\:\:\:\:\:\:\:\:\:\:\:\:\:\:\:\:\:\:\:\:\:\:\:\:\:\:\times (\bm x-\bm x')^2\:\text{ln}\left|\tfrac{1}{2}(\bm x-\bm x')^2\right|,\nonumber\\
    \mathcal{L}_{\omega}\!&= \!\lambda^2  \!\!\!\int\!\!\dd^3 \bm x \dd^3 \bm x' \!f(\bm x) f(\bm x') W_0(\bm x,\bm x')(\bm x - \bm x')^2.\label{Ls}
\end{align}
Notice that $P_0$ corresponds to the excitation probability of the detector if it were interacting with the vacuum of Minkowski spacetime. Eq. \eqref{PcsPflat} contains all corrections to the excitation probability of the detector up to second order in the detector size, as we have considered all terms of this order or lower in our computations. 

In Eq. \eqref{PcsPflat} we see corrections arising from five different fronts. The $M_{ij}\mathcal{Q}^{ij}$ term is associated with the spacetime volume element in the rest frame of the trajectory where the detector interacts with the quantum field. The $a_i\mathcal{D}^i$ term is the effect that the instantaneous acceleration of the detector has in the shape of its rest surface. The $R_{ij}\mathcal{L}^{ij}$ term is related to the corrections to the correlation function due to the Van-Vleck determinant, associated with the determinant of the parallel propagator. The $R \,\mathcal{L}_R$ term is associated to the corrections to the correlation function due to spacetime curvature. Finally, the $\omega_0 \,\mathcal{L}_\omega$ term is associated with the state of the quantum field. We highlight that this is the only term in Eq. \eqref{PcsPflat} whose coefficient is not independent of the other ones, given that we can write $\mathcal{L}_\omega = \delta_{ij}\mathcal{L}^{ij}$. 

The expansion of Eq. \eqref{PcsPflat} contains the effect of the curvature of spacetime in the excitation probability of a smeared delta-coupled UDW detector. Moreover, this expansion works for a large class of spacetimes under weak assumptions for the quantum field, provided that the detector size is small compared to the curvature radius of spacetime. It is also important to mention that the integral for $\mathcal{L}$ in Eq. \eqref{eq:L} is not solvable analytically in most spacetimes, and can demand great computational power to be performed numerically. However, the expression for $\mathcal{L}_0$, $\mathcal{Q}^{ij}$, $\mathcal{D}^i$, $\mathcal{L}^{ij}$, $\mathcal{L}_R$ and $\mathcal{L}_\omega$ can be computed analytically for a large class of smearing functions (see, for instance Appendix \ref{app:L-terms}). In this sense, the expansion presented in this section can be used to simplify the study of sufficiently small particle detectors in curved spacetimes.   

Overall, the expansion in Eq. \eqref{PcsPflat} shows the different ways that the background geometry manifests itself on ultra rapid localized measurements of a quantum field.  


\section{The curvature of spacetime in terms of the excitation probability}\label{sec:recovery}

In this section we will use the results of Section \ref{sec:curvature} in order to build a protocol by which one can obtain the curvature of spacetime from the excitation probability of delta-coupled particle detectors. In order to do so, we will consider explicit shapes for detectors, so that we can explicitly compute the $\mathcal{L}_0$, $\mathcal{Q}^{ij}$, $\mathcal{D}^{i}$, $\mathcal{L}^{ij}$,  $\mathcal{L}_R$ and $\mathcal{L}_{\omega}$ terms of Eq. \eqref{PcsPflat}, and obtain the curvature-dependent terms $M_{ij}$, $R_{ij}$ and $R$ from the excitation probabilities.

Before outlining the operational protocol which will allow us to recover the spacetime curvature, it is important to discuss the effect that the detector size has in the excitation probability in flat spacetimes. Consider a pointlike detector in Minkowski spacetime. After the ultra rapid coupling with the quantum field, this detector will be in a maximally mixed state, with excitation probability equal to $1/2$. The physical reason behind the detector ending up in a maximally mixed state is that it instantaneously probes all of the field modes. This generates a great amount of noise, which results in the detector state containing no information about the field. Overall, the size of the detector determines the smallest wavelength (largest energy modes) that it is sensitive to. Thus, increasing the size of the detector makes it sensitive to less energetic modes, which then decreases the excitation probability, according to Eq. \eqref{eq:P}. This allows one to obtain information about the field modes up to a cutoff determined by the inverse of the detector's size.

The discussion of the last paragraph can also be extended to curved spacetimes. In particular, the fact that a point-like detector delta-coupled to a quantum field ends up in a maximally mixed state also holds in general spacetimes. In fact, in the pointlike limit one ends up sampling smaller and smaller regions that are locally flat, and too small to be affected by curvature. This can be explicitly seen from Eq. \eqref{PcsPflat}, where all the correction terms are proportional to some power of the detector size (Eq. \eqref{Ls}). Similarly, as discussed in the case of flat spacetimes, a finite-sized particle detector will then couple to field modes of finite-sized wavelengths, and the effect of these modes in the particle detector will change its excitation probability. 
Moreover, the curvature in different directions will affect the modes that propagate in these directions differently. This implies that probing the quantum field with smeared delta-coupled particle detectors with different shapes should allow one to recover the spacetime curvature in different directions. 


We are now at a step where we can explicitly formulate a protocol where spacetime curvature can be recovered from ultra rapid local measurements of a quantum field. In order to do this, we will first have to make assumptions about the spacetime $\mathcal{M}$ and the events where we sample the field. As one would expect, in order to recover the classical curvature of spacetime in terms of expected values of quantum systems, one would require many samplings of the quantum field in similar conditions. Thus, we will require our spacetime to be locally stationary for the duration of the experiment\footnote{This is a strong condition that could be relaxed, as we only need spacetime not to vary too much in the frame of one timelike curve during the experiments, but we will assume this stronger version in order to build an explicit protocol.}, so that it contains a local timelike Killing field $\xi$ localized in the region where the experiments take place. Moreover, we will assume that the center points of the interactions of the particle detectors with the quantum field can all be connected by the flow of $\xi$. This will ensure that the curvature tensor $R_{\mu\nu\alpha\beta}$ and all other tensors derived from it are the same for all interactions considered, so that the expansion of Eq. \eqref{PcsPflat} has constant coefficients $M_{ij}$, $a_i$, $R_{ij}$ and $R$. 

The final assumption for our setup is that the different centers of the interactions are sufficiently separated in time so that the backreaction that each coupling of the detectors has on the field can dissipate away. This is a key assumption, which implies that the field state being probed remains approximately the same throughout the interaction. Equivalently, this implies that the state dependent part of the Wightman function expansion in Eq. \eqref{PcsPflat}, $\omega_0$, will remain approximately constant within the detectors smearings. We note that we are considering a massless field, so that field excitations propagate at light-speed. Thus, the assumption that $\omega_0$ is approximately constant translates into the different interactions being separated in time by more than the detectors' light-crossing time. Overall, this is a reasonable assumption for any experimental setup.

In order to build an explicit protocol, we will consider the detectors smearing functions to be given by ellipsoidal Gaussians in their respective Fermi normal frames. By considering ellipsoidal Gaussians as the shape of the detectors, we will then be able to select the modes that they are sensitive to in each spatial direction. Explicitly, we consider smearing functions of the form
\begin{equation}\label{eq:f}
    f(\bm x) = \frac{\sqrt{\det(a_{ij})}}{(2\pi)^\frac{3}{2}}e^{- \frac{1}{2}{a_{ij} x^i x^j}},
\end{equation}
where $a_{ij}$ is a positive symmetric bilinear map. We assume $\sqrt{\det(a_{ij})} = \mathcal{O}(L^{-3})$, where $L$ is a constant with units of length that determines the approximate size of the detectors and dictates the smallest wavelengths that they are sensitive to. The smearing function is prescribed in the detector's rest space in terms of the Fermi normal coordinates $\bm x = (x^1,x^2,x^3)$.

With the explicit choice of detectors shapes in Eq. \eqref{eq:f}, it is possible to compute most coefficients from the expansion of Eq. \eqref{PcsPflat} analytically. In fact, in Appendix \ref{app:L-terms}, we show that with the choice of elliptic Gaussian for $f(\bm x)$, $\mathcal{L}_0$, $\mathcal{Q}^{ij}$, $\mathcal{D}^i$, $\mathcal{L}^{ij}$ and $\mathcal{L}_\omega$ can be computed analytically. Moreover, $\mathcal{D}^i = 0$ in this case, so that the expansion in Eq. \eqref{PcsPflat} can be written as
\begin{equation}
    P = P_0 + e^{- 2 \mathcal{L}_0}\left(M_{ij}\mathcal{Q}^{ij}+N_{ij}\mathcal{L}^{ij} + \frac{2\pi^2}{3} R \,\mathcal{L}_R \right),\label{PP0}
\end{equation}
where $M_{ij} = \frac{2}{3} R_{\tau i \tau j} - \frac{1}{3}R_{ij}$ and $N_{ij} = \frac{1}{12}R_{ij} + 4\pi^2 \omega_0 \delta_{ij}$. In Appendix \ref{app:L-terms} we also show that $\mathcal{Q}^{ij}$, $\mathcal{L}^{ij}$ and $\mathcal{L}_R$ can all be varied independently due to their different non-linear dependence on $a_{ij}$ (or equivalently, on the shape of the detector). In this sense, Eq. \eqref{PP0} particularizes the expansion in Eq. \eqref{PcsPflat} for this specific setup and explicitly shows the independent coefficients $\mathcal{Q}^{ij}$, $\mathcal{L}^{ij}$ and $\mathcal{L}_R$ determined by the detectors' shape.

We are now at a stage where we can pick different detector sizes and shapes in order to recover information about the curvature of spacetime from their excitation probabilities. First, we consider the case where the detector's trajectory $\mf z(\tau)$ is the flow of the Killing vector field $\xi$. In this case, we expect to recover the tensors $M_{ij}$ and $N_{ij}$ and the scalar $R$ by sampling the probability $P$ in Eq. \eqref{PcsPflat} for different shapes of detectors (or, correspondingly, for different values of $a_{ij}$). That is, we perform measurements using different detectors with different shapes placed in different orientations, so that we ``sample the effect of curvature'' in each direction. In order to fully recover these tensors, it is necessary to sample the field using at least $13$ different values of $a_{ij}$ which give a set of $13$ \emph{linearly independent} coefficients $\mathcal{Q}^{ij}$ (with $6$ independent components), $\mathcal{L}^{ij}$ (with $6$ independent components) and $\mathcal{L}_R$. We then need a total of $13 = 6+6+1$ measurements in order to be able to write $M_{ij}$, $N_{ij}$ and $R$ in terms of the different probabilities. 

From the tensors $M_{ij}$, $N_{ij}$ and $R$ it is possible to recover $R_{ij}$, $R_{\tau i \tau j}$ and $\omega_0$. In fact, using \mbox{$M_{i}{}^i = \frac{2}{3} R_{\tau \tau} - \frac{1}{3}R_{i}{}^i$} and \mbox{$R = -R_{\tau\tau} + R_{i}{}^i$}, we obtain \mbox{$R_{i}{}^i = 2R + 3M_i{}^i$}. We can then obtain the state dependent term,  \mbox{$\omega_0 = \frac{1}{12\pi^2}\left(N_{i}{}^i - \frac{1}{12}R_{i}{}^i\right)$}. Finally, the curvature tensors can be written as \mbox{$R_{ij} = 12(N_{ij} - 4\pi^2 \omega_0 \delta_{ij})$} and \mbox{$R_{\tau i \tau j} = \frac{3}{2}M_{ij} + 2 R_{ij}$}. 
This protocol then allows one to recover $13$ independent terms: we recover all the space components of the Ricci scalar $R_{ij}$, all components of the Riemann tensor of the form $R_{\tau i \tau j}$ and the state dependent term $\omega_0$. In particular, from $R_{ij}$ and $R_{\tau i \tau j}$, it is possible to obtain $R_{\tau \tau}$ and the Ricci scalar $R$. The protocol outlined above then allows one to recover information about the spacetime geometry using only $13$ different couplings of detectors with the field. Moreover, if the spacetime whose geometry we wish to recover has known symmetries, it might be possible to require even less than $13$ samplings by exploiting these symmetries.


At this stage it should be clear that it is possible to recover some information about the spacetime geometry from the excitation probability of ultra rapid coupled particle detectors. However, it is still not possible to recover the full Ricci tensor, or the full Riemann curvature tensor from the setup described so far. In fact, it is not possible to write the components $R_{\tau i}$, $R_{\tau i j k}$ or $R_{ijkl}$ in terms of $M^{ij}$, $N^{ij}$ and $R$. However, it is possible to recover these tensors by considering detectors in different states of motion such that the center of their interactions with the quantum field is at events that still lie along the same flow of the Killing field $\xi$. For concreteness, consider a second detector which has a relative velocity $v$ in a (Fermi normal) coordinate direction $x^i$ with respect to the previous setup. In this case, it is possible to write the instantaneous four-velocity of the second detector at the point of the interaction as $\mf u' = \gamma(\mf u+v \mf e_i)$, where $\mf u$ is the four-velocity of the flow of the $\xi$ trajectory, $\mf e_i$ is the frame vector associated with the Fermi normal coordinates in the direction $i$ and $v$ is the magnitude of the instantaneous relative three-velocity between the trajectories. Then, performing the same protocol described above for detectors with relative velocity $v$ at the interaction points, we will obtain the tensors $R_{i'j'}$, $R_{\tau'i'\tau'j'}$ and the scalar $R_{\tau'\tau'}$, where the primed coordinates are associated with the components with respect to the Fermi frame of the trajectory $\mf u'$. Using the standard Lorentz coordinate transformation between these frames at the interaction points, it is possible to write \mbox{$R_{\tau\tau} = \gamma^2(R_{\tau'\tau'} - 2 v R_{\tau' i'} + v^2 R_{i'i'})$}. This expression now allows us to write the components $R_{\tau'i'}$ in terms of other previously obtained tensor components. An analogous procedure can also be carried over to the Riemann curvature tensor, allowing one to obtain $R_{\tau'i'j'k'}$ and $R_{i'j'k'l'}$ by considering frames with relative motion with respect to the flow of $\xi$. With this protocol, we are then able to recover all components of the Riemann curvature tensor.

We have particularized this protocol for specific elliptical gaussian detector shapes, so that their proper acceleration did not play any role in the expansion of Eq. \eqref{PcsPflat}. That is, this choice allows one to recover the geometry of spacetime regardless of the instantaneous proper acceleration of the detectors. However, it is possible to generalize this procedure using general detector shapes, provided that one finds linearly independent coefficients for the terms $\mathcal{Q}^{ij}$, $\mathcal{D}^i$, $\mathcal{L}^{ij}$ and $\mathcal{L}_R$. In fact, if we had considered detectors with nontrivial $\mathcal{D}^i$ terms, the acceleration of the detector would also play a role in the expansion of Eq. \eqref{PcsPflat}. Then, with $16$ couplings it would be possible to recover $M_{ij}$, $a_i$, $N_{ij}$ and $R$. An analogous protocol could then be performed in order to recover the full Riemann curvature tensor curvature of spacetime.

Overall, we have shown that it is possible to write the components of the curvature tensors in terms of the excitation probabilities of smeared delta coupled particle detectors of different shapes in different states of motion. In order to do so, we assume that the spacetime geometry is approximately unchanged for the duration of the experiments. Intuitively, by varying the shape of the detector in different directions, the detector will couple to different field modes, which are affected by curvature in specific ways according to Eq. \eqref{PcsPflat}. Having the specific dependence of these modes on curvature then allows one to associate the excitation probability of the particle detectors with the geometry of spacetime.

\section{Conclusion}\label{sec:conclusion}

We have expressed the spacetime curvature in terms of the excitation probability of smeared particle detectors delta coupled to a quantum field. Specifically, we devised a protocol in which one considers particle detectors of specific shapes and with specific states of motion which repeatedly interact with the quantum field. Under the assumption that the background geometry is approximately unchanged during these measurements, one can then recover the components of the Riemann curvature tensor associated with the directions in which each detector is more smeared. 

With the protocol we have devised, it is then possible to recover all components of the Riemann curvature tensor, and thus all information about the spacetime geometry, from measurable quantities of particle detectors. Overall, we have devised a protocol by which one can write the geometry of spacetime in terms of the expectation values of quantum observables. This represents yet another step towards obtaining a theory of spacetime and gravity which is compatible with with quantum theory and rephrases classical notions of spacetime and curvature entirely in terms of properties of quantum fields.


\section*{Acknowledgements}

The authors thank Bruno de S. L. Torres and Barbara \v{S}oda for insightful discussions and Erickson Tjoa for reviewing the manuscript. A.S. thanks Prof. Robert Mann for his supervision. T. R. P. thanks Profs. David Kubiz\v{n}\'ak and  Eduardo Mart\'in-Mart\'inez’s funding through their NSERC Discovery grants. Research at Perimeter Institute is supported in part by the Government of Canada through the Department of Innovation, Science and Industry Canada and by the Province of Ontario through the Ministry of Colleges and Universities.

\appendix

\section{The excitation probability in a quasifree state}\label{app:L}

We consider a quasifree state $\omega$ for a real scalar quantum field. The excitation probability for a delta-coupled detector according to Section \ref{sec:detectors} then reads
\begin{equation}
    P = \frac{1}{2}\left(1 - \omega(e^{2 i \hat{Y}})\right),
\end{equation}
where
\begin{align}
    e^{2 i \hat{Y}} = \sum_{n = 0}^\infty \frac{(2i\lambda)^n}{n!}\int \dd^3\bm x_1...\dd^3 \bm x_n &f(\bm x_1)...f(\bm x_n) \\
    &\:\:\:\:\:\times\hat{\phi}(\bm x_1)...\hat{\phi}(\bm x_n).\nonumber
\end{align}
In a quasifree state, we have $\omega(\hat{\phi}(\bm x)) = 0$, so that $\omega(\hat{\phi}(\bm x_1)...\hat{\phi}(\bm x_{2n+1})) = 0$ for all odd powers of the field. This allows one to write
\begin{align}\label{integrals}
    \omega(e^{2 i \hat{Y}}) \!= \sum_{n = 0}^\infty \frac{(-1)^n(2\lambda)^{2n}}{(2n)!}\!\!\!\int\! &\dd^3\bm x_1...\dd^3 \bm x_{2n} f(\bm x_1)...f(\bm x_{2n})\nonumber\\&\times \omega(\hat{\phi}(\bm x_1)...\hat{\phi}(\bm x_{2n})).
\end{align}
Moreover, we can use Wick's theorem in order to write the $2n$-point functions as
\begin{equation}
    \omega(\hat{\phi}(\bm x_1)...\hat{\phi}(\bm x_{2n})) = \sum_{\sigma \in P_{2n}} \prod_{i=1}^n W(\bm x_{\sigma(i)},\bm x_{\sigma(i+1)}),
\end{equation}
where we denote $W(\bm x,\bm x') = \omega(\hat{\phi}(\bm x)\hat{\phi}(\bm x'))$ and $P_{2n}$ denotes the set of $(2n-1)!!$ permutations involved in Wick's theorem.

At this stage, we notice that the functions of $2n$ variables $f(\bm x_1)...f(\bm x_{2n})$ are invariant under permutations. This implies that under a change of variables on the integrals of Eq. \eqref{integrals}, we obtain  that every term in the sum over the permutations in $P_{2n}$ yields the same result. That is, we obtain
\begin{align}
    \omega(e^{2 i \hat{Y}}) =& \sum_{n = 0}^\infty \frac{(-1)^n(2\lambda)^{2n}(2n-1)!!}{(2n)!}\\& \!\!\!\!\!\!\!\!\!\times\int \dd^3\bm x_1...\dd^3 \bm x_{2n} f(\bm x_1)...f(\bm x_{2n})\prod_{i=1}^n W(\bm x_{i},\bm x_{i+1}) .\nonumber
\end{align}
At this stage, we notice that the $2n$ integrals can be factored as $n$ identical integrals, given by $\mathcal{L}$ in Eq. \eqref{eq:L}. That is, we can write
\begin{equation}
    \omega(e^{2 i \hat{Y}}) = \sum_{n = 0}^\infty \frac{(-1)^n2^{2n}(2n-1)!!}{(2n)!}\mathcal{L}^n.
\end{equation}
Using $(2n-1)!! = 2^{1-n}(2n-1)!/(n-1)!$, we reach the simplified expression
\begin{equation}
    \omega(e^{2 i \hat{Y}}) = \sum_{n = 0}^\infty \frac{(-1)^n2^n}{n!} \mathcal{L}^n = e^{-2 \mathcal{L}},
\end{equation}
so that the excitation probability for a delta-coupled detector in flat spacetimes reads
\begin{equation}
    P = \frac{1}{2}\left(1 - e^{-2 \mathcal{L}}\right).
\end{equation}

\section{Fermi normal coordinates}\label{ap:fermi}

In this appendix we review some basic concepts associated to the Fermi normal coordinates (FNC) around a trajectory $\mf z(\tau)$ and expansions of the metric in this coordinate system. In order to build the FNC around a timelike curve $\mf z(\tau)$ with four-velocity $\mf u$, we pick a value $\tau_0$ and an orthonormal frame $\mf e_{\mu}(\tau_0)$ at $\mf z(\tau_0)$ such that $\mf e_0(\tau_0) = \mf  u(\tau_0)$ and $\mf e_{i}(\tau_0)$ satisfy $(e_i)_\mu u^\mu = 0$ and $(e_i)_\mu (e_j)^\mu = \delta_{ij}$. Then, we extend this frame to the whole curve $\mf z(\tau)$ by imposing that the $\mf e_{i}(\tau)$ are Fermi-Walker transported according to the equation
\begin{equation}
    \frac{\text{D}(e_i)^\mu}{\dd \tau} + a^{[\mu}u^{\nu]} (e_i)_\nu = 0.
\end{equation}
The Fermi-Walker transport above ensures that the extended vector fields $\mf e_i(\tau)$ remain orthogonal among themselves and to $\mf u$ for all $\tau$~\cite{poisson}.

With the frame $\mf e_i(\tau)$, we can write any vector $\mf v$ orthogonal to $\mf u$ at the point $\mf z(\tau)$ as $\mf v(\tau) = x^i \mf e_i(\tau)$. We then define the local rest space associated with $\mf z(\tau)$ at each $\tau$ as the set $\Sigma_\tau$, defined as the $\exp(\mf v(\tau))$ for every $\mf v(\tau)$ orthogonal to $\mf u(\tau)$ within the normal neighbourhood of $\mf z(\tau)$. $\Sigma_\tau$ also allows for a natural coordinate system built from this construction, in which we associate a point $p\in\Sigma_\tau$ to the coordinates $x^i$ if $\mf p = \exp(\mf v(\tau))$ with $\mf v(\tau) = x^i \mf e_i(\tau)$. This defines a coordinate system $x^i$ in each one of the local rest spaces of $\mf z(\tau)$. Thus, $(\tau,\bm x)$ define a coordinate system in a local region of spacetime defined by $\bigcup_{\tau \in \mathbb{R}}\Sigma_\tau$. This coordinate system is referred to as the Fermi normal coordinates associated to the curve $\mf z(\tau)$, and its applications range from the description of extended bodies in general relativity, to many computations involving local rest frames of observers~\cite{DixonI,DixonII,DixonIII,poisson,us,mine}.

In Fermi normal coordinates the spacetime metric can be expanded as
\begin{align}
    \bar{g}_{\tau\tau} &= -(1+2a_i {x}^i+(a_i {x}^i)^2+ R_{\tau i \tau j}{x}^ix^j) + \mathcal{O}(r^3),\nonumber\\
    \bar{g}_{\tau i} &= -\frac{2}{3} R_{\tau kij}{x}^k{x}^j + \mathcal{O}(r^3),\quad\\
    \bar{g}_{ij} &= \delta_{ij}-\frac{1}{3}R_{ikjl}{x}^k{x}^l + \mathcal{O}(r^3),\nonumber
\end{align}
where $r$ denotes the geodesic distance between the point $\mf x = (\tau,\bm x)$ and the trajectory $\mf z(\tau)$.

In particular, the metric determinant at each $\tau$ can be expanded as
\begin{align}
    \sqrt{-g} = 1 + a_i{x}^i +\frac{1}{2}M_{ij} {x}^i{x}^j+\mathcal{O}(r^3),
\end{align}
where
\begin{align}
    M_{ij} &= \left(R_{\tau i \tau j}-\frac{1}{3}(R_{1 i 1 j}+R_{2i2j}+R_{3i3j})\right),\nonumber\\
    &= \frac{2}{3}R_{\tau i \tau j} - \frac{1}{3} R_{ij}
\end{align}
and all tensors above are evaluated at the point $\mf z(\tau)$ so that they bear no dependence in $\bm x$.


\section{Explicit computation of the coefficients in the short distance expansion}\label{app:L-terms}

In this appendix we simplify the terms in Eq. \eqref{Ls}, in the case where the smearing function of the detector can be written as in Eq. \eqref{eq:f} in $(3+1)$ spacetime dimensions. In this case, most terms in Eq. \eqref{Ls} can be written as a single integral, which can be analytically computed for each choice of $a_{ij}$. In order to simplify the terms $\mathcal{M}^{ij}$, $\mathcal{L}^{ij}$, $\mathcal{L}_R$ and $\mathcal{L}_\omega$, define the functional
\begin{align}
    &\mathcal{B}[h] =\lambda^2 \int\dd^3 \bm x \dd^3 \bm x' f(\bm x) f(\bm x') W_0(\bm x,\bm x')h(\bm x,\bm x')\\ &= \frac{\lambda^2}{8\pi^2}\! \frac{\det(a_{ij})}{(2\pi)^3} \!\!\!\int\!\! \mathrm{d}^3\bm{x}\mathrm{d}^3\bm{x}'\! e^{-\frac{1}{2}a_{ij}x^jx^j}\!e^{-\frac{1}{2}a_{ij}x'^jx'^j}\!\!\frac{h(\bm{x},\bm{x}')}{(\bm{x}-\bm{x}')^2}.\nonumber
\end{align}
Notice that different choices for the function $h(\bm x,\bm x')$ give the different terms in Eq. \eqref{Ls}.

In order to simplify the integral in $\mathcal{B}[h]$, we first complete the square in the exponential using
\begin{equation}
    a_{ij}(x^jx^j + x'^ix'^j) = a_{ij}(x-x')^i(x-x')^j + 2 a_{ij}x^ix'{}^j.
\end{equation}
This allows us to rewrite the Gaussians in $\mathcal{B}[h]$ as 
\begin{equation}
    e^{-\frac{1}{2}a_{ij}x^jx^j}\!e^{-\frac{1}{2}a_{ij}x'^jx'^j} =  e^{-\frac{1}{2}a_{ij}(x-x')^i(x-x')^j}e^{-a_{ij}x^ix'^j}.
\end{equation}
We now consider the following change of variables
\begin{equation}
    \begin{cases}
        u^i&=\frac{1}{\sqrt{2}}(x^i-x'^i),\\
        v^i&=\frac{1}{\sqrt{2}}(x^i+x'^i).
    \end{cases}
\end{equation}
Using these new variables we can then write \mbox{$v^{(i}v^{j)} - u^{(i}u^{j)} = 2x^{(i}x'{}^{j)}$}, so that using the symmetry of $a_{ij}$, we obtain \mbox{$a_{ij}x^ix'^j = \frac{1}{2}a_{ij}(v^iv^j-u^iu^j)$}. This allows us to simplify $\mathcal{B}[h]$ to
\begin{equation}\label{huv}
    \mathcal{B}[h] \!=\! \frac{\lambda^2}{8\pi^2}\! \frac{\det(a_{ij})}{(2\pi)^3}\!\!\! \int\! \frac{\mathrm{d}^3\bm{u}}{2\bm u^2} e^{-\frac{1}{2}a_{ij}u^iu^j} \!\!\!\! \int\!\! \mathrm{d}^3\bm{v} e^{-\frac{1}{2}a_{ij}v^iv^j}\!h(\bm u,\bm v),
\end{equation}
where we denoted $h(\bm x,\bm x')$ expressed in terms of the new variables $\bm u$ and $\bm v$ by $h(\bm u,\bm v)$. 

In order to perform the integrals for different choices of $h(\bm u,\bm v)$, it will be useful to perform a spatial coordinate transformation that takes the coordinates $(x^1,x^2,x^3)$ in the rest frame of the detector to the principal axis of the ellipsoid described by $a_{ij}$. In these coordinates, the expressions for the tensors $\mathcal{L}^{ij}$ and $\mathcal{Q}^{ij}$ in Eq. \eqref{Ls} become simplified, as these tensors become diagonal. We denote the principal coordinates by primes, so that $(x^{1'},x^{2'},x^{3'})$ are aligned with the principal directions of $a_{ij}$ and $a_{i'j'}x^{i'}x^{j'} = a^2(x^{1'})^2 + b^2 (x^{2'})^2 + c^2 (x^{3'})^2$. With these conventions, the eigenvalues of $a_{ij}$ are $a^2, b^2$ and $c^2$, which control the inverse lengthscale of the detector in each of the principal axis. For convenience, we will assume that $a\leq b\leq c$. Using the principal axis of the ellipsoid that defines the detectors' shape, most integrals in Eq. \eqref{Ls} can be solved analytically in terms of the elliptical functions
\begin{align}
    E(\varphi;k) &= \int_0^\varphi \dd \theta\:\sqrt{1-k^2\sin^2\theta},\\
    F(\varphi;k) &= \int_0^\varphi \frac{\dd \theta}{\sqrt{1-k^2\sin^2\theta}}.
\end{align}

We start by computing the excitation probability of an elliptical Gaussian detector in flat spacetimes, $\mathcal{L}_0$. Picking $h(\bm u,\bm v) = 1$ in Eq. \eqref{huv} yields the $\mathcal{L}_0$ term. We find that
\begin{equation}
    \mathcal{L}_0 = \frac{\lambda^2}{16\sqrt{2}\pi^2}  \frac{\det(a_{ij})}{\sqrt{c^2-a^2}}F\left(\cos\!{}^{-1}\!\!\left(\frac{a}{c}\right);\frac{c^2-b^2}{c^2-a^2}\right).
\end{equation}
In the case of a spherical detector with characteristic size $\sigma$, that is, $a = b = c  = 1/\sigma$, we obtain
\begin{equation}
    \mathcal{L}_0 = \frac{\lambda^2}{16 \sqrt{2}\pi^2}\frac{1}{\sigma^2}.
\end{equation}
Notice that in the limit of a pointlike detector ($\sigma\rightarrow 0$), we obtain $\mathcal{L}_0\longrightarrow \infty$, so that the excitation probability in Eq. \eqref{eq:P} gives $1/2$, as expected.

The state dependent coefficient $\mathcal{L}_{\omega}$ can be obtained by picking $h(\bm u, \bm v) = 2 \bm u^2$, so that both integrals over $u$ and $v$ become the same, and one obtains a constant, which is independent of the specific shape of the ellipsoid that defines the detector,
\begin{equation}
    \mathcal{L}_\omega = \frac{\lambda^2}{8\pi^2}.
\end{equation}

The $\mathcal{D}^i$ term is obtained by choosing \mbox{$h(\bm u,\bm v) = \frac{1}{\sqrt{2}}(u^i +v^i)$}. We then obtain two integrals which can be shown to be $0$ due to parity:
\begin{align}\label{ZERO1}
    &\int \frac{\mathrm{d}^3\bm{u}}{2\bm u^2} e^{-\frac{1}{2}a_{ij}u^iu^j}  u^k = 0, 
    &\int \mathrm{d}^3\bm{v} e^{-\frac{1}{2}a_{ij}v^iv^j}v^k = 0.
\end{align}
This implies that for elliptic Gaussians, the $\mathcal{D}^i$ term does not contribute, and the acceleration of the curve does not play any role to leading order in the expansion of Eq. \eqref{PcsPflat}.

Choosing $h(\bm u,\bm v) = 2 u^i u^j$ in Eq. \eqref{huv} gives the $\mathcal{L}^{ij}$ term. It is diagonal in the principal basis associated to the ellipsoid's shape and its components are given by
\begin{widetext}
\begin{align}
    \mathcal{L}^{1'1'} &=- \frac{\lambda^2\sqrt{\det(a_{ij})}\sqrt{c^2-b^2}}{32\sqrt{2}\pi^2(b^2-a^2)}\left(\frac{E\left(\arcsin\left(\frac{b}{c}\right);\frac{c^2-a^2}{c^2-b^2}\right)}{{a^2-c^2}}+\frac{F\left(\arcsin\left(\frac{b}{c}\right);\frac{c^2-a^2}{c^2-b^2}\right)}{{c^2-b^2}}-\frac{2b}{ac}\right),\nonumber\\
   \mathcal{L}^{2'2'} &= \frac{\lambda^2\sqrt{\det(a_{ij})}\sqrt{c^2-a^2}}{16\sqrt{2}\pi^2}  \left(\frac{
   E\left(\arcsin\left(\frac{a}{c}\right);\frac{c^2-b^2}{c^2-a^2}\right)}{\left(b^2-a^2\right) \left(b^2-c^2\right)}-\frac{ F\left(\arcsin
  \left(\frac{a}{c}\right);\frac{c^2-b^2}{c^2-a^2}\right)}{(c^2-a^2) \left(c^2-b^2\right)}\right) + \mathcal{F}_2, \nonumber\\
    \mathcal{L}^{3'3'} &= \frac{\lambda^2\sqrt{\det(a_{ij})}}{16\sqrt{2}\pi^2}\frac{\left(-E\left(\arcsin\left(\frac{a}{c}\right);\frac{c^2-b^2}{c^2-a^2}\right)+F\left(\arcsin\left(\frac{a}{c}\right);\frac{c^2-b^2}{c^2-a^2}\right)\right)}{\sqrt{c^2-a^2} \left(c^2-b^2\right)},
\end{align}
where $\mathcal{F}_2$ is defined as
\begin{align}
    \mathcal{F}_2 &= \frac{\lambda^2}{16\sqrt{2}\pi^2}\frac{\sqrt{\det(a_{ij})}}{(2\pi)^{3/2}}\frac{a
   \left(a^2 \left(2 b^2+c^2\right)+c^2 \left(b^2-4 c^2\right)+6 c^3 \sqrt{c^2-b^2} \tanh ^{-1}\left(\sqrt{1-\frac{b^2}{c^2}}\right)\right)}{3b
   c^3 \left(a^2-b^2\right)^2}.
\end{align}
\end{widetext}

The term $\mathcal{Q}^{ij}$ can be obtained by choosing $h(\bm u,\bm v) = \frac{1}{2}(u^i + v^i)(u^j+v^j) = \frac{1}{2}(u^i u^j + v^i v^j + 2u^{(i}v^{j)})$. The integrals that involve mixed terms of $u$ and $v$ will be zero due to Eq. \eqref{ZERO1}. Also notice that the term corresponding to $\frac{1}{2}u^i u^j$ yields a result proportional to $\mathcal{L}^{ij}$, while the $\frac{1}{2}v^iv^j$ term gives a result proportional to $\mathcal{L}_0$, so that $\mathcal{Q}^{ij} = \frac{1}{4}\mathcal{L}^{ij} + \frac{1}{2}\mathcal{L}_0\mathcal{E}^{ij}$. Here $\mathcal{E}^{ij}$ is given by
\begin{equation}
    \mathcal{E}^{ij} =\left(\displaystyle{\int \dd^3 \bm v e^{-\frac{1}{2}a_{kl}v^kv^l}v^i v^j}\right)\!\!\bigg/\!\!\left(\displaystyle{\int \dd^3 \bm v e^{-\frac{1}{2}a_{kl}v^kv^l}}\right).\nonumber
\end{equation}
We note that $\mathcal{E}^{i'j'}$ is diagonal in the principal basis of $a_{ij}$. Its diagonal elements are given by
\begin{equation}
    \mathcal{E}^{1'1'} = \frac{\sqrt{2}}{a}, \qquad\mathcal{E}^{2'2'} = \frac{\sqrt{2}}{b},\qquad \mathcal{E}^{3'3'} = \frac{\sqrt{2}}{c}.
\end{equation}
Thus, $\mathcal{Q}^{ij}$ is also diagonal in this basis. Finally, notice that $\mathcal{Q}^{ij}$ can be varied independently of $\mathcal{L}^{ij}$ due to the nonlinear dependence of the $\mathcal{E}^{ij}$ term in $a_{ij}$.

Finally, we stress that the integral for $\mathcal{L}_R$ cannot be solved analytically for both $\bm u$ and $\bm v$ and a general $a_{ij}$. We can however write
\begin{equation}
    \mathcal{L}_R = \frac{\lambda^2}{8\pi^2}\frac{\sqrt{\det(a_{ij})}}{(2\pi)^{3/2}} \int \frac{\dd^3 \bm u}{2 \bm u^2} e^{-\frac{1}{2} a_{ij} u^i u^j} \ln( \bm u^2).
\end{equation}
Given that it is not possible to integrate the expression above analytically, this implies that $\mathcal{L}_R$ can be varied independently of all other coefficients computed above.




\bibliography{references}
\end{document}